\documentclass[12pt]{article}
\usepackage{graphicx}
\usepackage{graphics}
\usepackage{verbatim}   
\usepackage{color}      
\usepackage{subfigure}  
\usepackage{amssymb}    
\usepackage{amsmath}    
\usepackage{amsthm}     
\usepackage[dvips]{epsfig}

\usepackage{natbib}
\usepackage{lineno} 

\usepackage{float} 
\restylefloat{table}

\usepackage{booktabs} 
\newcommand{\ra}[1]{\renewcommand{\arraystretch}{#1}}

\graphicspath{{c:/Users/Roberto/Desktop/research/business/tourismpuertorico/writeup/figures/}}

\newcommand{\bolds}[1]{\mbox{\boldmath $#1$}} 

\bibpunct{(}{)}{;}{a}{}{,}

\begin{document}


\title{A Dynamic Linear Model to Forecast Hotel Registrations in Puerto Rico Using Google Trends Data}
\author{Roberto Rivera\footnote{University of Puerto Rico, Mayaguez}}

\maketitle

\pagestyle{plain} 


\section{Abstract}

Recently, studies have used search query volume (SQV) data to forecast a given process of interest. However, Google Trends SQV data comes from a periodic sample of queries. As a result, Google Trends data is different every week. We propose a Dynamic Linear Model that treats SQV data as a representation of an unobservable process. We apply our model to forecast the number of hotel nonresident registrations in Puerto Rico using SQV data downloaded in 11 different occasions. The model provides better inference on the association between the number of hotel nonresident registrations and SQV than using Google Trends data retrieved only on one occasion. Furthermore, our model results in more realistic prediction intervals of forecasts. However, compared to simpler models we only find evidence of better performance for our model when making forecasts on a horizon of over 6 months.

\section{Introduction}
In recent years, there has been an interest in exploiting search query data available through sources such as Google Trends (www.google.com/trends) to model temporal processes. \citet{choivarian12,choivarian09b} used search query data to model tourism demand, auto sales, home sales,
and initial unemployment claims. \citet{ginsbergetal09} relied on Google search queries to model influenza activity in the U.S. Studies have also suggested search query based tools to model consumer behavior \citep{goeletal2010}, dengue \citep{gluskinetal14} and more. Similarly, \citet{yangetal2014} used web traffic data to model hotel demand. It is not exactly known how the search query volume algorithm by Google generates its results. Moreover, the time series of search query volume generated by the algorithm changes every week. 

Puerto Rico has been going through an economic recession since 2006. Leaders on the island have been attempting to find ways to boost the economy. Although hotel registrations from July to November showed an increase of about 10\% from fiscal year 2012 to 2013 \citep{jp13a}, over the long term the contribution of the hotel industry to Gross Domestic Product has stayed relatively constant \citep{ruiz12}. With opportunities in many sectors of the economy dwindling, the government has been taking steps to improve the tourism sector. 
To accomplish this, efficient planning is crucial. Statistical  inference can be used to forecast the number of hotel registrations by nonresidents, a proxy of tourism demand. 

This is the first study to treat each weekly Google Trends output as a source of data of an unobservable process. We use this data to draw inference  on the lagged association between the number of hotel nonresident registrations (NHNR) in Puerto Rico and search query volume (SQV). The performance of our Dynamic Linear Model forecasting NHNR is compared to alternative models.


\section{Data}
\label{sec:methodstourism}
Number of hotel nonresident registrations from January 2004 to September 2012 was provided by the Puerto Rico Tourism Company, the local government agency that manages the industry on the island. Hotels and luxury hotels are required to provide registration data while short term stays and guest houses can provide it if they wish to do so.
Although NHNR does not exactly measure the number of tourists that come to the island, it intuitively serves as a good proxy.
A publicly available tool called Google Trends provides an index of relative volume of search queries based on a percentage of Google web searches.
The data quantifies the normalized volume of searches for a given query, typically over 7 days. We emphasize the use of the word `normalized' here, meaning the query volume at a given time point, divided by the maximum query volume of that term over the entire time series of interest. The resulting relative volumes are then scaled to a range of 0 to 100. Therefore the normalized SQV obtained are dependent on the region, category/subcategory, queries, and time frame selected. It should be noted that normalized search query volume is different than absolute search query volume. If the rate of absolute search query volume increase is smaller than the total search query volume, relative search query volume may decrease. The user may enter a search query or several search queries which Google Trends normalizes to provide SQV. To fit the model, the SQV Google Trends data, provided on a 7 day scale, was converted to monthly data. The possibility of using 7 day SQV data without converting to monthly data is discussed in section \ref{sec:conc_tourism}.

\subsection{Challenges in using Google SQV data} 
As appealing as the availability of the search query data is, care must be taken. \citet{butler13} found that Google Flu trends, a search query based tool, had not been performing as well as when it was introduced in 2009, sometimes estimating twice as many actual influenza cases. More recently, \citet{lazeretal14} showed that from August 21, 2011 to September 1, 2013, Google Flu Trends reported overly high flu prevalence 100 out of 108 weeks. Screening the search query data one finds that, for a fixed period of time of interest, fixed search queries and a fixed region, Google query output will differ over the time series. For example, if one is to obtain today data from Google Trends from 2004 to 2014 for ``puerto rico hotels" performed in the United States, one would obtain a time series of results. However, if one would extract output under the same settings next week, the time series has different entries. Every week the output will be different. This is different than data revision of economic data where only the most recent data changes. In the case of Google Trends, data at all time points change routinely. The issue is partly due to how the SQV data is provided through Google Trends. According to the help page of Google Trends, the companies algorithm analyzes a percentage of Google web searches to determine the amount of searches for the terms entered compared to the total number of Google searches done during the same time period. The statement implies that SQV is based on a sample of Google searches, but it doesn't specify the sample size or how samples are chosen. Another possible explanation for the poor performance of Google Flu Trends is the fact that Google constantly changes the functionality of searches \citep{lazeretal14}. Among recent changes, the use of social networking data and predicting misspellings to determine search results for users. Another aspect to consider is that Google constantly changes its search result ranking algorithm which may in principle affect the SQV time series. 
The challenges presented here are not to say that the search query data is not useful. One way of seeing it, is that the search query data provided by Google, is an observed version of the true search query process. 
\subsection{Choice of Google Trends Settings to obtain SQV data}
Search query volume data acquired from Google trends is a function of settings the user determines (e.g. region or location where searches were made, categories and subcategories of the search queries, the search type, etc.). The region to obtain the search volume data was chosen to be the United States. Including other countries would likely blur the association between Google SQV and NHNR for the following reasons. First, most nonresident tourists come from the United States. According to Puerto Rico's Tourism Company, for the 2011 fiscal year 92.6\% of visitors surveyed came from the United States  \citep{jp13b}. Of visitors arriving from the U.S. 44.1\% came from the East coast, most from New York and Florida. Secondly, Google's search market share overall is large, but it may vary considerably by country. Although no official numbers of search market share exists, estimates from several companies indicate that Google's market share is lower than local alternatives in some countries such as South Korea, China, Russia, and Japan. For such countries using search query volume data from other search engines may work best \citep{yangetal2015}. Lastly, exploration of queries related to Puerto Rico travel from countries other than the U.S. often produced little search volume data, sometimes no data at all.  
Table \ref{tab:trendssetting} summarizes the settings we used while using Google Trends.
\begin{table}[H]
	\begin{center}
		\begin{tabular}{|l||c|}\hline
			Search Volume Attribute & Chosen Setting\\\hline Queries &
			puerto rico hotels, puerto rico flights, san juan hotels,\\
			& puerto rico resorts, puerto rico vacations, puerto rico vacation, \\ & puerto rico tourism, puerto rico travel, and puerto rico hotel deals\\
			& \\
			Region & United States\\
			Search time frame & January 2004 - January 2014 \\
			Search type & Web Search \\ Category & Travel \\
			Subcategory & NONE\\
			Dates when volume data gathered & Every Thursday from 10/2/14 to 12/11/14 \\\hline\hline
		\end{tabular}\\
		\caption{Search Volume settings used with Google Trends to gather data to construct models to forecast the number of hotel nonresident registrations in Puerto Rico.}
		\label{tab:trendssetting}
	\end{center}
\end{table}
Using feedback from experts at the Puerto Rico Tourism Company and preliminary analysis, it was determined that the 9 queries shown in Table \ref{tab:trendssetting} where the best alternatives to forecast NHNR without exceeding the 30 word limit that Google Trends permits. The search query composite aggregating over all search queries allows a more complete time series (for separate search queries gaps may exist due to low volume), and avoids instability issues in parameter estimation of the model given that the search queries are highly correlated. Only Web search type volume was used from the Travel category. This Google Trends Travel category contains subcategories, but the search volume data for our queries of interest was spotty within these subcategories, so no subcategories were selected.

We used the results of these queries in the period from January 2004 to January 2014, but to fit the models we only used the time frame for which we have room registration data. Finally, SQV data was extracted in 11 consecutive Thursdays,  from October 2 to December 11, 2014. In the next section we discuss the models considered to study the capacity of query volume data to improve forecasts of NHNR.

\section{Forecasting Models}
We can express the NHNR data in a rather ambiguous form:
\begin{eqnarray}
	Y = g(\mu,S,\epsilon) \nonumber
\end{eqnarray}
That is, the data is decomposed into a trend or an association with search traffic component (modeled through $\mu$), seasonality component ($S$), and some irregular time dependence component ($\epsilon$). $g(\cdot)$ determines the type of function of these components. We model each component in an additive way based on stochastic approaches. Exploratory analysis gave no support to the use of non-parametric models such as artificial neural networks or regression splines \citep{htf09} for our data. 

\subsection{Dynamic Linear Model}
\label{sec:dlm}
The Dynamic Linear Model (DLM) is a flexible way to intuitively capture how processes evolve in time. 
In fact, traditional time series models such as ARIMA and others can be viewed as special cases of the DLM. Yet the dynamic linear model can also incorporate nonstationarity, time-varying parameters, multivariate time series, data from multiple sources, irregular temporal observations, and missing data among other things. \citet{shumwaystoffer11, chatfield03} provide nice introductions to DLMs while \citet{durbinkoopman12, brockwellanddavis09} cover more advance theory on the subject. DLMs have been widely used to model environmental data (\citet{cressiewikle11}, \citet{huertaetal04}), and economic or financial data \citet{shumwaystoffer11}. But the DLM has received much less attention in other business applications, and although it has been applied to model tourism data \citep{athanasopouloshyndman08, duPreezwitt03}, they have not been applied to Google Trends data as done in this study. Let $\bolds{Y}_{t}=(Y_{1,t},Y_{2,t},..., Y_{m,t})^{'}$ represent observations of $m$ time series at time $t$. Hence each $\bolds{Y}_{t}$ is a $m \times 1$ vector. Furthermore, let $\bolds{X}_{t}=(X_{1,t}, ..., X_{q,t})^{'}$ be the true $q$ processes of interest, and $\bolds{S}_{t}=(S_{1,t}, ...,S_{1,t-k}, ..., S_{q,t}, ...,S_{q,t-k})^{'}$ represents the seasonal component of period $k$ for each of the $q$ processes in the model. We express the DLM with the following equations:
\begin{eqnarray}
	\bolds{Y}_{t} = \bolds{F}\bolds{X}_{t} + \bolds{H}\bolds{S}_{t} + \bolds{\nu}_{t},
	\hspace{1cm} \bolds{\nu}_{t} \sim N(\bolds{0},\bolds{V})
	\label{eq:obsmodel}
\end{eqnarray}
\begin{eqnarray}
	\bolds{X}_{t} = \bolds{G}^{(x)}\bolds{X}_{t-1} + \bolds{C}\bolds{S}_{t} +
	\bolds{\omega}_{t}^{(x)},
	\hspace{1cm} \bolds{\omega}_{t}^{(x)} \sim N(\bolds{0},\bolds{W}_{t}^{(x)})
	\label{eq:evolmodel1}\\
	\bolds{S}_{t} = \bolds{G}^{(s)}\bolds{S}_{t-1} +
	\bolds{\omega}_{t}^{(s)},
	\hspace{1cm} \bolds{\omega}_{t}^{(s)} \sim N(\bolds{0},\bolds{W}^{(s)})\label{eq:evolmodel2}
\end{eqnarray}
Equation (\ref{eq:obsmodel}) is known as the observation or measurement equation, where $\bolds{\nu_{t}}$ corresponds to Normally distributed measurement error with mean zero
and covariance $\bolds{V}$. $\bolds{X}_{t}$ is referred to as the state or system vector \citep{westharrison97} and contains all the parameters that relate to
the trend
of the temporal processes of interest. The set of equations imply that the state vector of interest $\bolds{X}_{t}$ cannot be observed directly. $\bolds{F}$ is a $m \times q$ matrix that may depend on parameters that need to be estimated. $\bolds{H}$, $\bolds{C}$ are matrices with dimension and entries depending on whether the seasonal component is modeled as a fixed effect or stochastically (see section \ref{sec:ourdlm} for details on our approach).  \\ 

Equations (\ref{eq:evolmodel1}) and (\ref{eq:evolmodel2}) are known as state, system, or transition equations. These equations determine how $\bolds{X}_{t}$ is generated from past values $\bolds{X}_{t-1}$. $\bolds{G}_{t}^{(s)}$ and $\bolds{G}_{t}^{(x)}$ are referred to as the evolution matrices with
dimensions $s \times s$ and $q \times q$ respectively,  and $\bolds{\omega}_{t}^{(s)}$, $\bolds{\omega}_{t}^{(x)}$ are the evolution errors. As stated in \citet{bcg03}, usually the design problem at hand
determines the form of $\bolds{F}$ while modeling assumptions
lead to how $\bolds{G}^{(s)}, \bolds{G}^{(x)}$ are represented. Specifically, dependence among $(Y_{1,t},Y_{2,t},...Y_{m,t})^{'}$ can be introduced into the model through $\bolds{G}^{(s)}, \bolds{G}^{(x)}$ or $\bolds{W}_{t}^{(x)}, \bolds{W}^{(s)}, \bolds{V}$. Choosing the identity matrix as $\bolds{G}^{(x)}$ results in a random walk
representation for $\bolds{X}_{t}$ for all $t$. More generally $\bolds{F}, \bolds{H}, \bolds{G}^{(s)}, \bolds{G}^{(x)}$ may be time
dependent sequence of matrices, an extension that we do not pursue here.

\subsubsection{Dynamic Linear Model for NHNR} \label{sec:ourdlm}
To adapt the
DLM to our case let $\bolds{Y}_{t}=(Y_{1,t},Y_{2,t},..., Y_{(a+1),t})^{'}$ where $Y_{1,t}$ is the recorded number of hotel nonresident registrations for time $t$ and $Y_{2,t},..., Y_{(a+1),t}$ are query volume data retrieved from Google Trends from their algorithm runs $1, ... ,a$ for time $t$. Hence each $\bolds{Y}_{t}$ is a $(a+1) \times 1$ vector. Of prime importance to us is the true process $\bolds{X}_{t}=(X_{1,t},X_{2,t})^{'}$ where  $X_{1,t}$ is the true NHNR for time $t$ and $X_{2,t}$ is the true SQV data for time $t$. Assuming a seasonal component of period $k=12$, which we model as a fixed factor, our DLM consists of the following equations in matrix form:
\begin{eqnarray}
	\begin{pmatrix}
		Y_{1,t} \\
		\vdots \\
		Y_{(a+1),t} 
	\end{pmatrix} = \begin{pmatrix}
	1 & 0 \\
	0 & 1 \\
	\vdots & \vdots \\
	0 & 1 
\end{pmatrix}\begin{pmatrix}
X_{1,t} \\
X_{2,t} 
\end{pmatrix} + \begin{pmatrix}
\nu_{1,t} \\
\vdots \\
\nu_{(a+1),t} 
\end{pmatrix}
\label{eq:ourobseqn}
\end{eqnarray}
where the observation errors are assumed to be independent from the state vector for all t, and no correlation is assumed between the observation errors $\nu_{1,t}$ and $\nu_{j,t}, j=2,...,(a+1)$. Conversely, correlation between  $\nu_{j,t}, j=2,...,(a+1)$ are possible. However, not constraining the correlation between all $\nu_{j,t}, j=2,...,(a+1)$ requires the estimation of too many\footnote{In our data $a=11$, hence not constraining the correlation between all $\nu_{j,t}, j=2,...,12$ requires the estimation of $(121-11)/2=55$ off-diagonal parameters in $\bolds{V}$. Moreover, a fixed covariance among all 10 search query output errors did not improve the model.} off-diagonal parameters in $\bolds{V}$.  
Therefore we assume $\bolds{V} =diag(\sigma^{2(y)}_{1}, \sigma^{2(y)}_{2}I)$ where $I$ is an identity matrix. 
The other equation looks as follows,
\begin{eqnarray}
	\begin{pmatrix}
		X_{1,t} \\
		X_{2,t} 
	\end{pmatrix} = \begin{pmatrix}
	1 & \beta \\
	0 & 1
\end{pmatrix}\begin{pmatrix}
X_{1,t-1} \\
X_{2,t-1} 
\end{pmatrix}  + \bolds{C}\bolds{S}_{t} + \begin{pmatrix}
\omega_{1,t} ^{(x)}\\
\omega_{2,t}^{(x)} 
\end{pmatrix}
\label{eq:ourevolutioneqn1}
\end{eqnarray}

where variances $\bolds{W}_{t}^{(x)} = diag(\sigma^{2(x)}_{1}, \sigma^{2(x)}_{2})$, and the errors $\{\bolds{\nu}_{t}\}, \{\bolds{\omega}_{t}^{(x)}\}$ are uncorrelated. 
Note that (\ref{eq:ourobseqn}) and (\ref{eq:ourevolutioneqn1}) imply that the Seasonal component for each temporal process is modeled as fixed with $\bolds{H}=\bolds{0}$, $\bolds{S}_{t}$ indicating the month at time $t$, $\bolds{C}$ is a $2 \times 12$ matrix of parameters, and no stochastic component ($\bolds{W}^{(s)} = diag(\sigma^{2(s)}_{1}, \sigma^{2(s)}_{2})=0$).
The $\beta$ parameter is linked to the linear association between NHNR at time $t$ and SQV at time $t-1$. $\beta \neq 0$ would indicate that there is a linear association between $X_{2,t-1}$ and $X_{1,t}$ while $\beta=0$ would indicate no linear association between these processes and hence, no practical use of SQV in forecasting NHNR. Also, $\beta \neq 0$ implies that the true SQV is a leading indicator of NHNR. Leading indicators are useful in forecasting processes of interest, since they don't have to be forecasted themselves for short lead times. In this case, however, $X_{2,t-1}$ is not directly observed making its usefulness as a leading indicator less clear. This DLM allows us to account for the information from multiple Google Trends algorithm runs to determine if there is a linear association between $X_{2,t-1}$ and $X_{1,t}$. 
We hypothesized that incorporating data from multiple search volume algorithms had an impact in determining the type of association between the true NHNR and the true SQV. We tested this hypothesis by comparing the inference on $\beta$ from the DLM expressed above and a DLM using only the most recent search query algorithm data.

\subsubsection{Estimation of parameters and Kalman recursions}\label{sec:ourdlminf}
Parameters in equations (\ref{eq:ourobseqn}), (\ref{eq:ourevolutioneqn1}) must be estimated. Direct Maximum likelihood estimation methods \citep{brockwellanddavis09}, Expected Maximization (EM), and Bayesian methods \citep{westharrison97} are some alternatives. In this work we used the EM algorithm described in \citep{holmes12} and implemented through the \textbf{\textsf{R}} \citep{R15} using package MARSS \citep{holmesetal12}. Briefly, the EM algorithm allows for maximum-likelihood estimation in models with unobserved latent variables. The method works through iterations between two steps. First step is the expectation of the log likelihood with respect to the unobserved variables given observed data and current parameter estimates. In the second step, the expected log likelihood is maximized to obtain new parameter estimates. Iterations continue until the log-likelihood is maximized. MARSS uses a Monte Carlo algorithm to choose starting values for the EM algorithm that will result in the highest likelihood. Confidence intervals for $\beta$ and other parameters were based on asymptotic Normality and an estimated Hessian matrix. 
Predictions $\hat{X}_{i,t^{'}}$ and $\hat{Y}_{i,t^{'}}$ at times $t^{'}$ for $i=1,2$ were obtained using Kalman Recursions \citep{brockwellanddavis09} through the DLM presented in this paper.

\subsection{Benchmark forecasting models}\label{sec:othermodels}
A Seasonal Autoregressive Integrated Moving Average (SARIMA), Holt-Winter (HW), and a seasonal na\"{\i}ve (SNAIVE) model were also fit. \citet{shumwaystoffer11} discusses the SARIMA model while \citet{chatfield03} briefly explains Holt-Winter models. SARIMA and Holt-Winter have been used in the past to model tourism arrivals and they tend to perform well \citep{duPreezwitt03, limmcaleer01}. For a review of recent tourism demand modeling approaches see \citet{songetal13} and \citet{songli08}. The SNAIVE will forecast NHNR based on the most recent month. Hence the forecast for June 2013 is the NHNR of June 2012, showing that SNAIVE is the least technical model of our alternatives. For theoretical relationships of dynamic linear models with ARIMA models, Holt-Winter and other methods see \citet{durbinkoopman12}. For this study, all models were fitted using data from January 2004 to September 2012. Data from October 2012 to September 2013 was held out to compare the forecasting performance of all models.

\section{Results}
\begin{figure}[h]
	\begin{center}
		\includegraphics[scale=0.5]{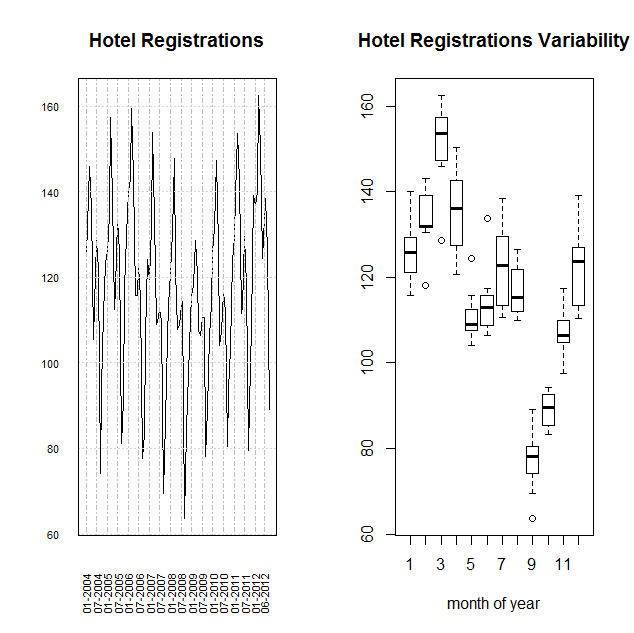}\\

	\end{center}
	\caption{Left panel displays the Time series plot of nonresident hotel registrations in Puerto Rico (in thousands). Right panel shows boxplots of nonresident hotel registrations (in thousands) summaries by month of year.} \label{fig:timeseriesandboxplot}
\end{figure}

Over the broader time period from January 2004 to December 2012, peak number of hotel nonresident registrations occurred in 2012 (1,575,131), while 2010 and 2011 had an increase slightly above 5\% from the previous year (1,305,532 for 2009, and 1,373,786 for 2010). However, from 2005 to 2009, a yearly decrease in NHNR occurred. The lowest registrations occurred in 2009, likely related to the financial crisis in the U.S. As expected, the hotel room registration data displayed a strong seasonal pattern (left panel Figure \ref{fig:timeseriesandboxplot}). Highest NHNR occurred around the dry season months (December to April) with a peak in March while lowest NHNR occurred in the wet months with September providing the lowest occupancy (right panel Figure \ref{fig:timeseriesandboxplot}). In fact, seasonality dominates the time series, suggesting that the Holt-Winter's model is a viable option to generate forecasts of NHNR. The seasonality did not appear to vary widely on a year to year basis. 

\begin{figure}[h]
	\begin{center}
		\includegraphics[scale=0.5]{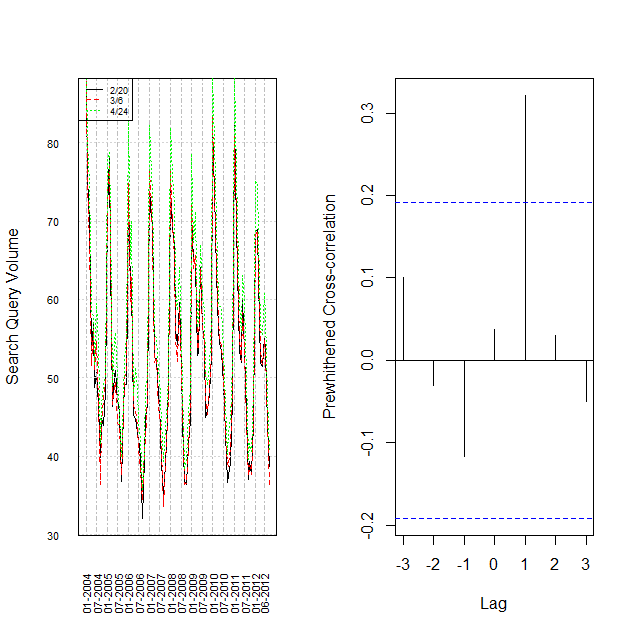}\\

	\end{center}
	\caption{Output from 3 separate search query volume output can be seen in the left panel. Cross correlation based on prewhitened nonresident hotel registrations and prewhitened search query volume (averaged over the 11 algorithmic outputs) are presented in the right panel.} \label{fig:timeseriesandcross}
\end{figure}

Turning to the search query volume data, the left panel of Figure \ref{fig:timeseriesandcross} shows the time series obtained  October 9, October 23 and December 11 of 2014. Although each time series was similar, variability among algorithm dates are visible. December 11 output had more pronounced seasonal peaks than the other Google Trends output, especially later in the time series. On the other hand October 9 and October 23 output displayed lower seasonal bottoms than December 11 output. When omitting seasonality neither of the two temporal processes displayed major changes in average value over time nor a strong increasing or decreasing trend. A Canova-Hansen test on the NHNR and SQV data did not reject the null hypothesis of deterministic seasonality. Also, an Augmented Dickey-Fuller test applied to the times series rejected the presence of a unit root. Based on the sample autocorrelation function and the partial autocorrelation function and initial number of autoregressive, moving average, seasonal autoregressive, and seasonal moving average parameters was chosen. The final number of parameters for the SARIMA model was determined by minimizing the Akaike Information Criteria, also known as AIC \citep{akaike73}. Ljung-Box test applied to model residuals showed no remaining autocorrelation. Time series were prewhitened as suggested in \citet{bisgaardkulahci11} to inspect cross correlation. As we can see from Figure \ref{fig:timeseriesandcross} (right panel), there appears to be a significant one month lag association between the NHNR and SQV time series. However, the lag-1 cross correlation was estimated to be 0.32, implying an association of only moderate strength.  
%

\subsection{DLM results to determine association between $X_{1,t}$ and $X_{2,t-1}$}
\begin{figure}[h]
	\begin{center}
		\includegraphics[scale=0.4]{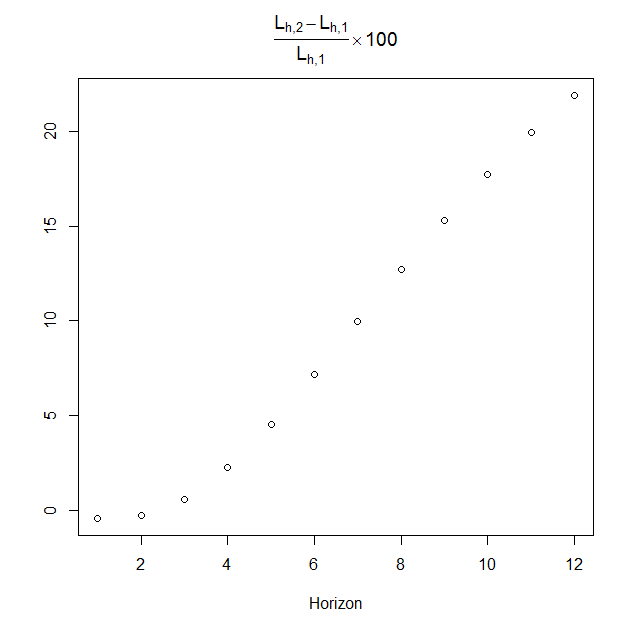}\\

	\end{center}
	\caption{Percentage difference in length of $DLM_{2}$ relative to $DLM_{1}$ as a function of forecast horizon. $L_{h,i}=$ upper bound - lower bound of the $DLM_i$ prediction interval for horizon $h$. Overall the $DLM_2$ prediction intervals are too optimistic, especially for larger forecast horizons.}\label{fig:dlm1dlm2PIcomparisonbyhorizon}
\end{figure}

We fitted two DLMs as presented in sections \ref{sec:ourdlm} and \ref{sec:ourdlminf} to draw inference on $\beta$. The multivariate time series $\bolds{Y}_{t}$ was demeaned before constructing the DLM models. For one model, $DLM_{1}$, we used Google Trends generated time series of search query volumes for $a_{1}=11$ weeks. The second DLM model, $DLM_{2}$, used only the most recent Google Trends search query volume data $a_{2}=1$. Table \ref{tab:predvaravgestimates} shows the resulting estimates of $\beta$, 95\% confidence intervals based on asymptotic Normality and forecast accuracy measures mean absolute error (MAE), and mean absolute percentage error (MAPE) using one step ahead forecasts. We see that the estimate of $\beta$ through $DLM_{2}$ was virtually the same than through $DLM_{1}$. However, the $\beta$ confidence interval based on $DLM_{1}$ did not include zero while the one based on $DLM_{2}$ did. Furthermore, $DLM_{1}$ had smaller length than the confidence interval based on $DLM_{2}$. The length of a confidence interval is the difference between its upper and its lower bound. By using output of 11 Google trends algorithm runs we can better infer about the linear association between $X_{1,t}$ and $X_{2,t-1}$. No difference was detected on the inference drawn from both models regarding $\sigma^{2(y)}_{1}$ and $\sigma^{2(x)}_{1}$. But, the results on $\sigma^{2(y)}_{2}$ and $\sigma^{2(x)}_{2}$ from $DLM_{1}$ imply that the search query volume is on average evolving in time while the results of $DLM_{2}$ put this in doubt with a confidence interval lower bound closer to zero. Furthermore, while we only have a slight decrease in MAE and MAPE when using $DLM_{1}$, the Kalman forecast variance equation is a function of $\sigma^{2(x)}_{2}$ \citep[see]{shumwaystoffer11}. By treating search query volume as an unobservable process, $DLM_{1}$ is able to give a more realistic estimate of $\sigma^{2(x)}_{2}$ than $DLM_{2}$. Therefore, since $DLM_{2}$ estimate of $\sigma^{2(x)}_{2}$ is almost 3 times smaller than the $DLM_{1}$ estimate, $DLM_{2}$ prediction intervals will generally be too optimistic (smaller in length than what they should be). The recursive nature of the Kalman forecasts will make the difference in length of the $DLM_{1}$ and $DLM_{2}$ prediction intervals greater as the forecast horizon increases. Figure \ref{fig:dlm1dlm2PIcomparisonbyhorizon} presents the percentage difference in length of $DLM_{2}$ relative to $DLM_{1}$ as the forecast horizon increases. For a 12 month ahead forecast, $DLM_{2}$ results on a prediction interval that is about 22\% smaller than the $DLM_{1}$ prediction interval. 

\begin{table*}[h]
	\centering
	\ra{1.3}
	\resizebox{\columnwidth}{!}{%
		\begin{tabular}{@{}lcccccccc@{}}\toprule
			& \multicolumn{5}{c}{Parameter estimates}  & \phantom{abc} & \multicolumn{2}{c}{Forecast accuracy}\\
			\cmidrule{1-5} \cmidrule{6-9}
			Model & $\beta$ & $\sigma^{2(y)}_{1}$ & $\sigma^{2(y)}_{2}$ & $\sigma^{2(x)}_{1}$ & $\sigma^{2(x)}_{2}$ && $MAE$ & $MAPE$\\ \midrule
			$DLM_{1}$ & 104.56 & $1.25\times 10^{7}$  & 1.63  & $2.89\times 10^{6}$  & 13.66  && 3560.78 & 3.18\\
			& (2.6, 206.52) & ($8.25\times 10^{6}$, $1.76\times 10^{7}$) & (1.50, 1.78) & ($8.70\times 10^{5}$, $6.10\times 10^{6}$) &  (10.11, 17.74) &  && \\
			\\
			$DLM_{2}$ & 104.72  & $1.26\times 10^{7}$ & 5.04 & $2.87\times 10^{6}$ & 4.68 && 3599.80 & 3.21\\
			& (-13.03, 222.47) & ($8.26\times 10^{6}$, $1.78\times 10^{7}$) & (2.65, 8.19) & ($8.08\times 10^{5}$, $6.21\times 10^{6}$) & (2.03, 8.44) &  && \\
			\bottomrule
		\end{tabular} %
	}
	\caption{Comparison of the inference on parameters and forecast accuracy using $DLM_{1}$ and $DLM_{2}$. 95\% confidence intervals (in parenthesis) were based on asymptotic Normality. The 95 \% confidence interval for $DLM_{2}$ implied no linear association between $X_{1,t}$ and $X_{2,t-1}$ while the one for $DLM_{1}$ implied a statistically significant association.}
	\label{tab:predvaravgestimates}
\end{table*}

The inference on $\beta$ supports the preliminary argument made in the previous section, suggesting that although a statistically significant linear association exists between $X_{1,t}$ and $X_{2,t-1}$, this association appears to be moderate. In the next section we compare the performance of our $DLM_{1}$ with the models outlined in section \ref{sec:othermodels}.


\subsection{Forecast accuracy}
\begin{figure}[h]
	\begin{center}
		\includegraphics[scale=0.5]{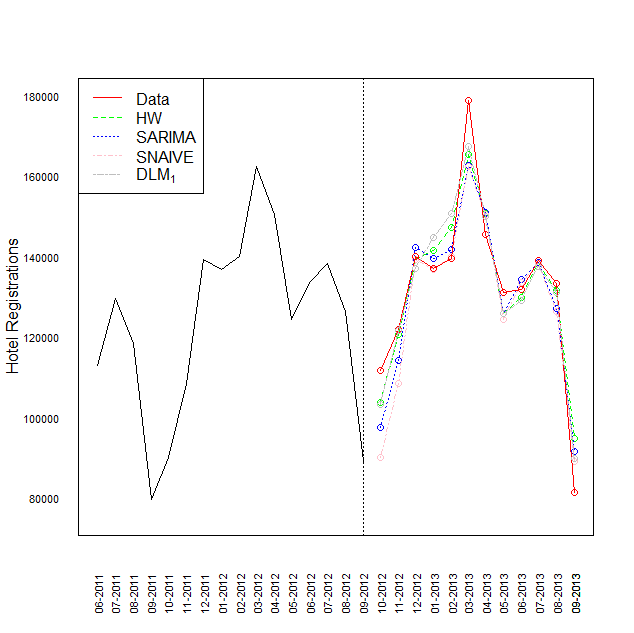}\\

	\end{center}
	\caption{Last few nonresident hotel registrations with forecasts up to 12 months ahead from all the models. Data not used to construct the model is also included.} \label{fig:roomdatawithmanymodelforecasts}
\end{figure}

The most common methods to determine forecasting accuracy are functions of forecasting error.  MAE, MAPE, and root mean square prediction error (RMSE) were calculated in sample, and out of sample for a time horizon of 6 months, and 7-12 months ahead. Given the small amount of out of sample data, statistical inference on significance of difference in forecasting errors is unreliable and not presented here. Summaries of the errors are presented in Table \ref{tab:prederrors}. A dynamic linear model without using SQV, $DLM_{0}$, was also fit for this comparison to take into account the modeling procedure while comparing forecasting performance. Based on the in sample results, SNAIVE had the worst fit to the data followed by SARIMA. Out of sample errors indicate that no model performed best in terms of forecasting over the short and long horizon simultaneously. Generally, prediction errors suggested HW was the best alternative for short term forecasts. For horizons over 6 months, $DLM_{1}$ performed best and neither SARIMA, HW, or $DLM_{0}$ performed convincingly better than SNAIVE. As we can see from Figure \ref{fig:roomdatawithmanymodelforecasts}, the forecast of all the models captured the overall pattern in NHNR, though all forecasts underestimated the March 2013 NHNR (which turned out to be higher than in any other March) and overestimated the September 2013 NHNR.  
\begin{table*}[h]
	\centering
	\ra{1.3}
	\begin{tabular}{@{}lccccccccccc@{}}\toprule
		& \multicolumn{10}{c}{\textbf{Forecast accuracy}}& \phantom{abc}\\\hline
		& \multicolumn{3}{c}{\bolds{$MAE$}} & \phantom{abc} & \multicolumn{3}{c}{\bolds{$MAPE$}} & \phantom{abc} & \multicolumn{3}{c}{\bolds{$RMSE$}}\\
		\cmidrule{2-4} \cmidrule{6-8} \cmidrule{10-12}
		\textbf{Model} & \textbf{In} & \textbf{Out-6} & \textbf{Out-12} && \textbf{In} & \textbf{Out-6} & \textbf{Out-12} && \textbf{In} & \textbf{Out-6} & \textbf{Out-12} \\ \midrule
		$DLM_{1}$ & 3560.78 & 7024.38 & 4160.76 && 3.18 & 5.01 & 3.76 && 4570.37 & 8111.41 & 4760.00\\
		$DLM_{0}$ & 3633.70 & 6485.74 & 4490.72 && 3.26 & 4.56 & 4.13 && 4653.87 & 7751.16 & 5282.41\\
		$SARIMA$ & 4709.43 & 7491.45 & 4972.53 && 4.14 & 5.50 & 4.50 && 6021.21 & 9451.35 & 5829.13\\
		$HW$ & 4220.69 & 5901.93  & 4756.51 && 3.80 & 4.14 & 4.59 && 5422.68 & 7330.08 & 6364.35\\
		$SNAIVE$ & 6702.11 & 8822.67 & 4762.50 && 5.72 & 6.77 & 4.14 && 8213.00 & 12373.1 & 5437.31\\
		\bottomrule
	\end{tabular}
	\caption{Forecast accuracy comparison of models. `In' column shows in sample errors, `Out-6' errors up to 6 months ahead and `Out-12' forecast errors for horizons of 7-12 months ahead. The out of sample period is October, 2012 to September, 2013.}
	\label{tab:prederrors}
\end{table*}

\begin{figure}[h]
	\begin{center}
		\includegraphics[scale=0.5]{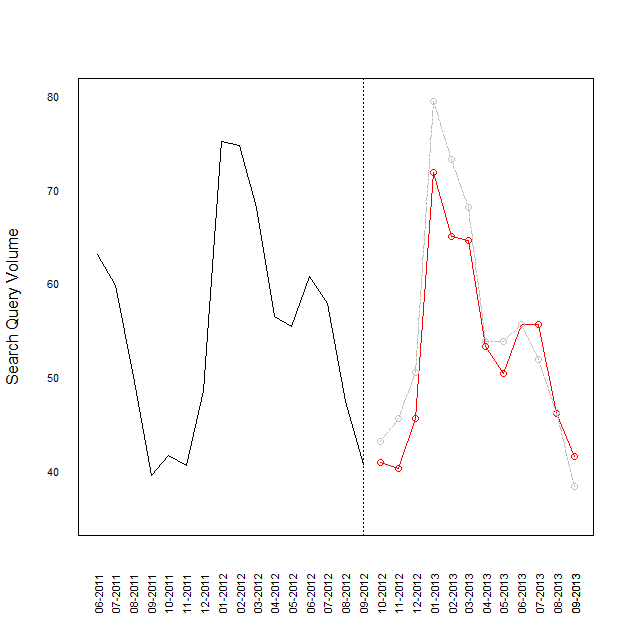}\\

	\end{center}
	\caption{Last few search query data observations (based on Google Trends algorithm run 11) with $DLM_{1}$ forecasts up to 12 months ahead (dotted line). Data not used to construct the model is also included (solid line).} \label{fig:searchquerywithDLMforecasts}
\end{figure}

Figure \ref{fig:searchquerywithDLMforecasts} presents the last few search query data observations (based on Google Trends algorithm run 11) with $DLM_{1}$ forecasts up to 12 months ahead. Data not used to construct the model is also included. We see that the model tended to overestimate the monthly search query volume for the first few months. Over the first 6 forecast months the MAPE when forecasting SQV was found to be 9.67 and for the forecasts 7-12 months ahead the MAPE was 3.74. The prediction errors over the first 6 forecast months are markedly higher than those for NHNR. Since the $DLM_{1}$ forecasts of $X_{1,t}$ depend on the forecasts of $X_{2,t-1}$, a poor performance in forecasting the latter process will hinder its accuracy in forecasting the former \citep{ashley83}. Models with autoregressive features and with a growth component were also considered for $X_{2,t}$ but they did not improve the results seen here.

\section{Conclusions}\label{sec:conc_tourism}
The aim of this work was to test the performance of a model using SQV data with other alternatives when forecasting the number of hotel nonresident registrations in Puerto Rico. As far as we know, this is the first paper to account for the uncertainty of the Google SQV data. We showed that our proposed DLM allows to conduct more precise inference on the lagged linear association of the two temporal processes than downloading Google Trends output only once. The evidence showed a statistically significant linear association between $X_{1,t}$ and $X_{2,t-1}$. Also the prediction intervals from our proposed model were more realistic than a model using just one sample output of the SQV time series. However, DLM forecast performance was mixed: for the shorter forecast horizon Holt-Winter works best, while for the longer forecast horizon $DLM_{1}$ works best. Three explanations are given for the forecasting performance of our DLM. First, the weakly moderate linear association between $X_{1,t}$ and $X_{2,t-1}$ inhibits $DLM_{1}$ from providing good forecast universally over short and long forecast horizons. \citet{duPreezwitt03} obtain similar findings where univariate models outperformed multivariate ones due to the absence of strong cross correlation between the processes. Secondly, the performance of the DLM in forecasting $X_{2,t}$ for the shorter forecast horizon was not good enough to compensate for the weakly moderate linear association between the processes. Third, over the long forecast horizon, $DLM_{1}$ is the only alternative that universally beats the seasonal na\"{\i}ve model, suggesting that there is value in using SQV data but only for long term forecasts of NHNR in Puerto Rico. The Holt-Winter model does not require expertise to fit, as the SARIMA and DLM do. Moreover, since the Holt-Winter resulted in the smallest forecast errors for horizons of 6 months or less, it is a more useful alternative to managers in the tourism sector for this forecast horizon. Long term forecasts may be used to determine policies, budget and other decisions related to the tourism sector, and we have shown that at longer horizons our DLM model using SQV performs better than simpler candidates. Moreover, since the SNAIVE only accounts for seasonal dependence to produce forecasts, our results indicate that Holt-Winter and SARIMA models should not be used for long term forecasts. 

We acknowledge that our selection of SQV data was mostly heuristic. Further research is needed using more objective alternatives to choose which and how many search queries should be included within the limits that Google Trends allows. Moreover, the association between NHNR and SQV may be stronger at a weekly level, since some visitors may schedule their stay a few weeks before making their trip instead of a month before hand. More research is needed to see if a dynamic model incorporating a latent process, and mixed frequency time series data would help improve forecasts. \citet{BangwayoSkeete2015} used a mixed frequency approach to forecast tourism arrivals in the Caribbean using search query data. However, our preliminary analysis indicates that the SQV data at a weekly level was noisier than its aggregated monthly counterpart, leading to higher prediction errors when forecasting SQV. At the very least, a stochastic seasonal component would be needed but the preliminary analysis does not support pursuing a mixed frequency model in our case. SQV data retrieved from Google Trends may improve short term forecasts of processes, especially in situations when the main time series of interest and the SQV data display strong growth and when the search query volume data can be forecast well. Yet, the SQV data is relative, not absolute and its absolute SQV that makes most sense of being associated with NHNR. Google does not provide much detail on how they obtain their SQV data and why the data available through Google Trends changes routinely. The mechanism producing the data helps determine the right modeling approach (e.g determining if there's a need to adjust for bias). More transparency from Google would improve the chances of exploiting the promising tool of search query volume data.


Care must be taken when analyzing the forecasting accuracy of models. To account for the uncertainty of prediction error statistics, hypothesis test methods based on forecast accuracy measures have been developed. However, simulations suggests that these methods require a substantial amount of out of sample data, at least 40 observations in length to be useful \citep{ashley03}. Also, although Kalman recursion allows for the estimate of prediction error covariance recursively, this estimate is dependent on assumptions taken about the covariance of the observation and system error.
Typically, in practice the covariances parameters are unknown and must be estimated. A fully Bayesian perspective would allow to measure the uncertainty involved in the estimation of these covariance parameters, an issue to be explored in the future.

\section*{Acknowledgments}
\thispagestyle{empty}
The author would like to thank the Puerto Rico Tourism Company for providing data and input especially Rafael Silvestrini. We are also grateful to the anonymous referees for their helpful recommendations.

\clearpage



\bibliographystyle{c:/Users/Roberto/Desktop/research/localtexfiles/bibtex/bst/misc/asa} 
\bibliographystyle{asa}
\bibliography{c:/Users/Roberto/Desktop/research/spatial/FRK/bibtexreferences2011}       

\end{document}